\documentclass{article}
\usepackage{waspaa19,amsmath,graphicx,url,times}
\usepackage{color}
\usepackage{amssymb}
\usepackage{booktabs}

\title{Model-agnostic Approaches to Handling Noisy Labels \\ When Training Sound Event Classifiers}

\name{Eduardo Fonseca,\sthanks{This work is partially supported by the European Union's Horizon 2020 research and innovation programme under grant agreement No 688382 AudioCommons and a Google Faculty Research Award 2018.}
      Frederic Font,
      Xavier Serra}
\address{Music Technology Group, Universitat Pompeu Fabra, Barcelona \{name.surname\}@upf.edu\\
}

\begin{document}

\ninept
\maketitle

\begin{sloppy}

\begin{abstract}
Label noise is emerging as a pressing issue in sound event classification.
This arises as we move towards larger datasets that are difficult to annotate manually, but it is even more severe if datasets are collected automatically from online repositories, where labels are inferred through automated heuristics applied to the audio content or metadata.
While learning from noisy labels has been an active area of research in computer vision, it has received little attention in sound event classification.
Most recent computer vision approaches against label noise are relatively complex, requiring complex networks or extra data resources.
In this work, we evaluate simple and efficient model-agnostic approaches to handling noisy labels when training sound event classifiers, namely label smoothing regularization, mixup and noise-robust loss functions.
The main advantage of these methods is that they can be easily incorporated to existing deep learning pipelines without need for network modifications or extra resources.
We report results from experiments conducted with the FSDnoisy18k dataset.
We show that these simple methods can be effective in mitigating the effect of label noise, providing up to 2.5\% of accuracy boost when incorporated to two different CNNs, while requiring minimal intervention and computational overhead.
\end{abstract}


\vspace{-1mm}
\section{Introduction}
\label{sec:intro}
\vspace{-1mm}
Deep neural networks require large and varied data resources in order to show their superior performance with respect to traditional machine learning methods, a fact that has become evident in fields like computer vision.
In the less explored field of sound event recognition, we are currently moving from small and exhaustively labeled datasets of few hours of duration \cite{salamon2014dataset,piczak2015esc,foster2015chime}, towards larger datasets in the range of tens (e.g., FSDKaggle2018 \cite{fonseca2018general} or FSDnoisy18k \cite{Fonseca2019learning}) to thousands (e.g., AudioSet \cite{gemmeke2017audio}) of hours of audio.
The increasing size of datasets makes it hard to manually label the audio content reliably as it turns out to be difficult and costly. 
This inevitably incurs in a certain amount of label noise either due to incomplete or incorrect annotations, even if produced by trained humans.

Online repositories such as Freesound or Youtube host significant volumes of audio content with associated metadata that can be used to create audio datasets. 
Labels can be inferred automatically by applying automated heuristics to the metadata, or pre-trained classifiers on the audio content.
While this way of collecting labeled data is much faster than the conventional dataset creation, the level of label noise generated can be much more severe.
Hence, label noise is a problem in large-scale sound event classification that can hinder the proper learning of classifiers, especially if they are based on deep neural networks~\cite{arpit2017closer,zhang2016understanding}.

The topic of \textit{learning with noisy labels} is an active area of research in computer vision. 
The state-of-the-art is based on selecting the clean data instances in the train set in order to train the network satisfactorily \cite{han2018co,malach2017decoupling,jiang2017mentornet}.
However, those methods can turn relatively complex as they leverage two networks (sometimes trained simultaneously).
Other methods rely on estimating the noise transition matrix, i.e., the probability of each true label being flipped into another \cite{patrini2017making,goldberger2016training}.
However, such estimation is not trivial, and it assumes that the only possible type of noise is flipping labels. 
Other approaches use noise-robust loss functions to mitigate the effect of label noise \cite{zhang2018generalized}, or leverage an additional set of curated data, for example to train a label cleaning network in order to reduce the noise of a dataset \cite{veit2017learning}.
Conversely, learning with noisy labels has received little attention in sound recognition, probably given the traditional paradigm of learning from relatively small and exhaustively labeled (hence \textit{clean}) datasets.
In our previous work, the FSDnoisy18k dataset is presented along with an evaluation of noise-robust loss functions \cite{Fonseca2019learning}. 
In \cite{kumar2018learning}, two networks operating on different views of the data co-teach each other to learn from noisy labels.
Recently, the topic of label noise was included for the first time as one of the research problems in the DCASE2019 Challenge~\cite{Fonseca2019audio}.

Most of the aforementioned approaches against label noise require complex networks (often more than one) or extra data resources.
Given the early stage of this field in sound event classification, we are interested in exploring simple and efficient approaches, agnostic to network architecture, that can mitigate the effect of label noise.
Specifically, we seek approaches that can be plugged into existing learning settings composed by a noisy dataset and a deep network, without need for network modifications or extra resources.
Our contribution is to provide insight on the model-agnostic approaches that can be incorporated to deep learning pipelines for sound event classification in presence of noisy labels, as well as the performance boost that can be expected.
In particular, we consider regularization techniques external to the model, as well as noise-robust loss functions (Fig. \ref{fig:idea}).
Regularization aims to prevent overfitting and improve generalization, which can also be beneficial against label noise.
Common regularization strategies include weight decay and dropout \cite{srivastava2014dropout}, which act on the weights or hidden units of the network; dropout has been shown useful in reducing label noise memorization \cite{arpit2017closer}.
In our attempt to regularize the model from the outside, we consider label smoothing regularization (LSR) and \textit{mixup}.
The former operates on the ground truth labels, while the latter operates on both ground truth labels and input data (Fig. \ref{fig:idea}).
In addition, we explore two strategies to ignore potentially noisy instances in the learning process through noise-robust loss functions.
Section \ref{sec:methods} describes the methods considered.
Section \ref{sec:setup} introduces the experimental setup.
Section \ref{sec:experiments} describes the experiments carried out and the results.
Section~\ref{sec:conclusion} provides final remarks.
\begin{figure}[t]
  \vspace{-2mm}
  \centering
  \centerline{\includegraphics[width=\columnwidth]{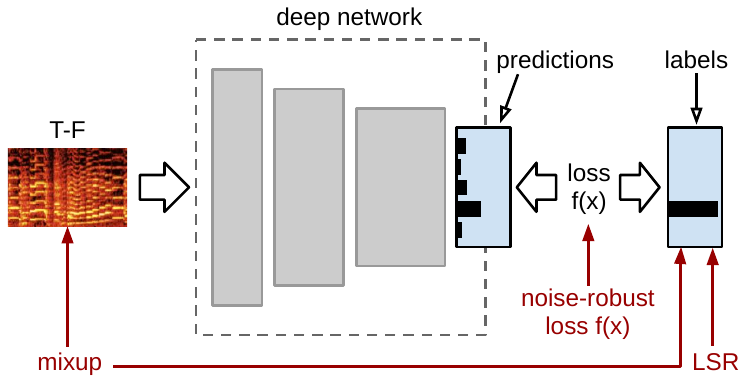}}
  \vspace{-2mm}
  \caption{Sketch of the model-agnostic approaches against label noise considered (in red), indicating the component(s) of the learning pipeline where they operate.}
  \vspace{-5mm}
  \label{fig:idea}
\end{figure}
\vspace{-3mm}

\vspace{-5mm}
\section{Methods}
\label{sec:methods}
Consider a dataset $\mathcal{D}$ with training examples $(x,y)$ and $K$ classes. 
Let $y(k)$ be a vector representing the ground truth distribution over labels with $k \in \left\lbrace 1 ... K \right\rbrace$.
We consider a multi-class classification problem (i.e., only a single ground truth label is available per example).
Hence, the ground truth distribution is a one-hot encoded vector $y(k)=\delta_{k,t}$, with $\delta_{k,t}$ being the Dirac delta, which equals 1 for the target label, $k = t$, and 0 otherwise.
Assume categorical cross-entropy (CCE) as default loss function, given by
\vspace{-2mm}
\begin{equation}
\label{eqn:cce}
\mathcal{L}_{cce}=-\sum_{k=1}^Ky(k)\log(p(k)),
\vspace{-2mm}
\end{equation}
where $p(k)$ represent the network softmax predictions.
The training goal is to update the network weights in order to minimize CCE, which means to maximize the log-likelihood of the correct label.

\vspace{-2mm}
\subsection{Label Smoothing Regularization}
\label{ssec:lsr}
Maximizing the log-likelihood of the (potentially) correct label means encouraging the model to be confident about its predictions, which can be harmful in presence of training data with noisy labels.
To address this issue, label smoothing regularization (LSR) models the noise on the labels using a smoothing parameter $\varepsilon$, as explained next \cite{szegedy2016rethinking}.
Given a training instance, the ground truth label distribution is changed from a one-hot encoded vector $y(k)=\delta_{k,t}$, to:
\vspace{-1mm}
\begin{equation}
y'(k) = (1-\varepsilon)\delta_{k,t}+\frac\varepsilon{K}.
\label{eqn:lsr}
\end{equation}
This is equivalent to the combination of the original distribution $y(k)$ weighted by $1 - \varepsilon$, and another distribution (in this case, uniform, i.e., $1/K$) weighted by $\varepsilon$. 
Essentially, we \textit{smooth} the label distribution by replacing the 1 and 0 hard classification targets with float soft targets.
Other distributions can be used to spread $\varepsilon$ across the non-active labels, e.g., using prior knowledge of the dataset.
By using LSR we are not seeking to maximize the log-likelihood of training labels, but to regularize the model by promoting less confident output distributions without discouraging correct classification \cite{goodfellow2016deep}.
This property makes the model less vulnerable to label noise.

\vspace{-2mm}
\subsection{mixup}
\label{ssec:mixup}
Mixup is typically understood as a data augmentation technique that acts as regularizer by favoring linear behavior in-between training examples \cite{zhang2017mixup}.
Specifically, mixup augments the training distribution by creating virtual examples under the assumption that linear interpolations in the feature space correspond to linear interpolations in the label space. This is expressed by 
\vspace{-1mm}
\begin{equation}
\begin{split} 
\tilde{x} = \lambda x_{i}+(1-\lambda)x_{j}\\
\tilde{y} = \lambda y_{i}+(1-\lambda)y_{j},
\end{split} 
\label{eqn:mixup}
\end{equation}
where $x_{i}$ and $x_{j}$ are two spectrograms from the training data, and $y_{i}$ and $y_{j}$ their corresponding one-hot encoded vectors.
As proposed in \cite{zhang2017mixup}, we sample $\lambda$ from a beta distribution $\lambda \sim$ Beta$(\alpha, \alpha)$, for $\alpha \in (0, \infty)$.
Since $\lambda \in [0, 1]$, we have convex combinations of the input spectrograms, with the hyper-parameter $\alpha$ controlling the strength of the interpolation.
Zhang et al. report that mixup enhances the robustness of deep networks to corrupted labels for image recognition in an experiment inflicting artificial label noise to a clean dataset \cite{zhang2017mixup}.
In this paper, we evaluate mixup on real-world label noise for sound event classification.
Note that mixup produces soft targets for the new training instances, in a similar fashion to LSR. However, mixup produces examples with only two active labels (two-hot encoded vectors) whereas with LSR no label is strictly inactive due to the smoothing distribution of $\varepsilon$ (see (\ref{eqn:lsr})).
In principle, mixup seems a reasonable regularization/augmentation strategy for sound events (even more than for image classification) as it roughly simulates the general setting of two sound sources $(x_{i}, x_{j})$ in an acoustic scene. 
Each of them will have a certain saliency, i.e., a sound pressure level (controlled by $\lambda$ in mixup), due to the attenuation produced by their different source-microphone distance.

\vspace{-2mm}
\subsection{Noise-Robust Loss Functions}
\label{ssec:losses}
In the process of minimizing a loss function, the weights' update can be suboptimal if the ground truth labels are corrupted, thereby hindering model convergence.
In these cases, loss functions that are robust against label noise can be helpful.
CCE is the loss function commonly used in multi-class classification tasks; however, CCE is sensitive to label noise as it puts more emphasis on \textit{hard} or \textit{difficult} examples. 
Due to the logarithm in (\ref{eqn:cce}), the examples for which the softmax predictions differ more from the ground truth labels are also weighed more in the gradient update.
This weighting property is beneficial when dealing with clean data, but it can be undesirable in the case of noisy labels. 
In contrast, the mean absolute error equation (MAE) treats every instance equally (hence avoiding the weighting aspect), and it has been shown theoretically that MAE can be used as a loss function robust against label noise \cite{ghosh2017robust}.
Nevertheless, MAE has been reported to cause other difficulties in training that lead to performance drop \cite{Fonseca2019learning,zhang2018generalized}.
To take advantage of the benefits of CCE (its weighting property) and MAE (its noise-robustness), a generalization of those functions has been recently proposed in~\cite{zhang2018generalized}.
This noise-robust loss function, termed $\mathcal{L}_{q}$, is the negative Box-Cox transformation of the softmax predictions:
\vspace{-1.5mm}
\begin{equation}
\label{eqn:zhang}
\mathcal{L}_{q}=\frac{1-(\sum_{k=1}^K y(k) p(k))^q}{q}, \quad q \in (0,1],
\vspace{-1.5mm}
\end{equation}
where $q$ controls the closeness of $\mathcal{L}_{q}$ to CCE or MAE.
$\mathcal{L}_{q}$ is the top performing noise-robust loss function among several evaluated for sound event classification \cite{Fonseca2019learning}. 
Further details about $\mathcal{L}_{q}$ can be found in~\cite{zhang2018generalized}.
Arpit et al. showed that deep neural networks in presence of label noise tend to learn first easy and general patterns from the underlying clean data, before fitting or memorizing the noise \cite{arpit2017closer}.
In other words, the negative impact of label noise becomes more severe as learning progresses.
This motivates us to view the learning process as a two-stage process.
In the \textit{first stage}, we adopt $\mathcal{L}_{q}$ during a training period of $n_1$ epochs.
While in the first epochs it is likely that label noise is not extremely critical (as long as there are some clean and simple instances in the dataset), it is not trivial to know when noise memorization kicks in as this depends on the data and the model; hence it is safer to adopt a noise-robust loss function.
After $n_1$ epochs, the model has converged to some extent and therefore it can be used for instance selection during training, a concept used previously in the label noise literature \cite{han2018co,zhang2018generalized}.
Thus, in this \textit{second stage}, we use the current state of the model to identify instances with large training loss.
Assuming they correspond to noisy labeled examples, the goal is to ignore them for the gradient update, thereby reducing noise memorization.

We experiment with two approaches to ignore large loss instances in the second stage.
The first one consists of \textit{discarding} large loss instances from each mini-batch of data dynamically at every iteration. 
This is based on a time-dependent, noise-robust loss function that exhibits a change of behaviour as learning progresses, starting to discard potentially corrupted instances after $n_1$ epochs.
The second approach consists of \textit{pruning} the train set after $n_1$ epochs based on loss values, keeping only a subset of the train set to continue the learning process.
To this end, we use the current model checkpoint to make predictions on the entire train set, and we compute the $\mathcal{L}_{q}$ losses associated to the softmax predictions.
Regardless the approach (discarding from each mini-batch, or pruning the train set), the rejection of large loss examples is as follows.
Once we have an array $\boldsymbol{\mathcal{L}_{q}} \in \mathbb{R}^{N\times1}$, with the loss for every example in the mini-batch (N=64) or the train set (N=13,441 after keeping 15\% for validation), we define a threshold $t_m$ such that elements in $\boldsymbol{\mathcal{L}_{q}}$ greater than $t_m$ are rejected.
We experiment with two simple ways of defining $t_m$: \textit{i)} $t_m = m\cdot max(\boldsymbol{\mathcal{L}_{q}})$ with $m \in [0,1]$, and \textit{ii)} $t_m = percentile(\boldsymbol{\mathcal{L}_{q}}, l)$ where $l$ is the percentile $\in [0,100]$.




\vspace{-2mm}
\section{Experimental Setup}
\label{sec:setup}
\vspace{-1mm}

The evaluation of the model-agnostic methods is conducted using the FSDnoisy18k dataset \cite{Fonseca2019learning}, an open dataset containing 42.5 hours of audio across 20 sound event classes, including a small amount of manually-labeled data and a larger quantity of real-world noisy data.
The audio content is taken from Freesound \cite{font2013freesound}, and the dataset was curated using the \emph{Freesound Annotator} \cite{Fonseca2017freesound}.\footnote{\url{https://annotator.freesound.org}}
We use only the noisy set of FSDnoisy18k, composed of 15,813 audio clips (38.8h), and the test set, composed of 947 audio clips (1.4h) with correct labels.
The dataset features two main types of label noise: \emph{in-vocabulary} (IV) and \emph{out-of-vocabulary} (OOV).
IV applies when, given an observed label that is incorrect or incomplete, the true or 
missing label is part of our target class set. 
Analogously, OOV means that the true or missing label is not covered by those 20 classes.
Further details can be found in \cite{Fonseca2019learning}.
The dataset's baseline system is a CNN of 532k weights composed of three convolutional and one fully connected layers with pre-activation \cite{fonseca2018simple}.
The loss function is CCE, the batch size is 64, and we use Adam optimizer \cite{kingma2014adam} with initial learning rate of 0.001, halved whenever the validation accuracy plateaus for 5 epochs.
Earlystopping is adopted with 15 epochs patience on the validation accuracy.
To this end, a 15\% validation set is split randomly from the training data of every class.
Input audio is transformed to 96-band, log-mel spectrogram, using time-frequency (T-F) patches of 2s.
We evaluate the approaches described in Section~\ref{sec:methods} by incorporating them to the baseline system.
Furthermore, the strategies with noise-robust loss functions are tested with a model of higher capacity.
We use a CNN based on Dense Convolutional Networks (DenseNet) \cite{huang2017densely}, which improve information flow in the network by connecting layers directly to all subsequent layers, combining their features by concatenation.
In particular, we use four \textit{dense} blocks composed of a bottleneck layer and a convolutional layer.
In addition, we include Squeeze-and-Excitation (\textit{SE}) blocks that calibrate the features extracted channel-wise by modelling channel interdependencies \cite{hu2018squeeze}.
These architectural blocks have been shown useful for sound event classification \cite{Jeong2018}.
In this work, we refer to this model as \textit{DenSE} due to its composition.
DenSE is more accurate than the baseline (see Table~\ref{tab:results_loss}) while using less weights (458k).
Further information about the CNN architectures used is available in the code release, along with the code for the experiments.\footnote{\url{https://github.com/edufonseca/waspaa19}\label{repo}}


\vspace{-3mm}
\section{Experiments}
\label{sec:experiments}
\begin{table}[!t]
\vspace{-3mm}
\caption{Average classification accuracy (\%) and 95\% confidence interval (7 runs) obtained by LSR and mixup approaches incorporated to the baseline system.}
\vspace{+1mm}
\centering
\begin{tabular}{lc}
\toprule
\textbf{Approach} & \textbf{Accuracy} \\
\midrule
Baseline \cite{Fonseca2019learning}	    & 66.5 $\pm$ 0.6	         \\
\midrule
LSR ($\varepsilon = 0.1$) 	                & 66.8 $\pm$ 1.0		   \\
LSR ($\varepsilon = 0.15$)	                & 67.1 $\pm$ 1.1	      \\
LSR ($\varepsilon = 0.15 \pm 0.05$)	        & 68.1 $\pm$ 0.8	      \\
\midrule
mixup ($\alpha = 0.1$)                      & 67.1 $\pm$ 0.8         \\
mixup ($\alpha = 0.2$)                      & 66.6 $\pm$ 0.7         \\
warm-up (10 epochs) \& mixup ($\alpha = 0.3$)      & 68.4 $\pm$ 0.5          \\
\bottomrule
\end{tabular}
\label{tab:results}
\vspace{-5mm}
\end{table}

\vspace{-1mm}
\subsection{Label Smoothing Regularization}
\label{ssec:ex_lsr}
First, the default version of LSR as described in Section \ref{ssec:lsr} is evaluated for several $\varepsilon$, which implies a uniform distribution of $\varepsilon$ across the non-active labels. 
This can also be seen from a probabilistic perspective, where the observed label is correct with probability $1 - \varepsilon$, and otherwise, any other label can be correct with equal probability \cite{goodfellow2016deep}.
Results in Table~\ref{tab:results} indicate a small improvement over the baseline system.
Larger values of $\varepsilon$ do not lead to better scores in our experiments.
In addition, we experiment with different smoothing strategies leveraging prior knowledge of FSDnoisy18k.
A per-class estimation of the label noise is available at the dataset companion site.
Based on this information, we group the audio categories in two groups, according to the estimated amount of noise (low/high).
Specifically, the low-noise group is composed by the categories \emph{Bass guitar}, \emph{Clapping}, \emph{Crash cymbal}, \emph{Engine}, \emph{Fire}, \emph{Rain}, \emph{Slam}, \emph{Walk, footsteps} and \emph{Wind}, while the high-noise group is complementary.
Then, we assign a different $\varepsilon$ to each group such that $\varepsilon_{low} = \varepsilon-\Delta \varepsilon$ and $\varepsilon_{high} = \varepsilon+\Delta \varepsilon$, where we grid-search for $\Delta \varepsilon \in \left\lbrace 0.025, 0.05\right\rbrace$, the latter providing best results.
This simple way of encoding prior knowledge of label noise through a noise-dependent $\varepsilon$ leads to the best LSR-based performance.
However, a finer grouping of the categories in three levels of noise (low/mid/high) does not provide further gain.
We also experiment with non-uniform smoothing distributions in order to model per-class information, in particular: \textit{i)} mapping each estimated amount of per-class noise to a per-class label energy within the ground truth vector, and \textit{ii)} a distribution based on the number of per-class T-F patches in the dataset. 
However, adding this level of specificity leads to performance degradation.

\vspace{-3mm}
\subsection{mixup}
\label{ssec:ex_mixup}
We conduct experiments with the default version as explained in Section \ref{ssec:mixup} for several values of the interpolation strength $\alpha$.
Examples to be mixed up are log-mel patches drawn randomly from the training data. 
In particular, we try both intra- and inter-batch variants, that is, applying mixup to examples of the same batch after random permutation, or to examples of two different batches. 
No major differences are observed.
In \cite{zhang2017mixup}, the authors carry out experiments using mixup against memorization of corrupted labels and find out that larger values of $\alpha$ (e.g., $\left\lbrace 8,32 \right\rbrace$) perform best.
They hypothesize that increasing $\alpha$ creates virtual examples further away from the training distribution, thus hampering noise memorization.
Surprisingly, in our experiments we find that larger values of $\alpha$ do not yield any improvement, the best accuracy being obtained with $\alpha = 0.1$, as seen in Table~\ref{tab:results}.
The main differences between our work and \cite{zhang2017mixup} are that FSDnoisy18k features real-world label noise (mainly of OOV type) in sound events, while \cite{zhang2017mixup} considers artificial label noise of IV type (randomly flipping labels) in images.

We also evaluate mixup by using a \textit{warm-up} training period in which mixup is not applied.  
This is motivated by experiments conducted also in \cite{zhang2017mixup}, in this case for speech recognition, although unrelated to label noise mitigation.
We choose warm-up periods of 5 and 10 epochs, and we evaluate $\alpha \in \left\lbrace 0.1, 0.2,0.3,0.4,1,2\right\rbrace$.
Warm-up based mixup shows a significant improvement over its default version, as can be seen in Table~\ref{tab:results}, the highest accuracy being obtained with a warm-up period of 10 epochs and $\alpha = 0.3$.
A possible explanation for the effectiveness of mixup is that continuously creating different virtual examples hinders label noise memorization.
Also, it could contribute to reduce the overall exposure to label noise. 
For example, if we consider two training examples inputting mixup, the low range of $\alpha$ used means that one input example clearly dominates over the other in the virtual example (due to the properties of the beta distribution).
Whenever only one input example is correctly labeled, we would occasionally move from learning from one correct and another incorrect label, to learning from one almost correct label and another one which is not entirely wrong.
A more in-depth analysis would be required to better understand how mixup mitigates the effect of label noise.

\vspace{-2mm}
\subsection{Noise-Robust Loss Functions}
\label{ssec:ex_losses}
We carry out experiments substituting the CCE loss with the proposed noise-robust loss function approaches described in Section~\ref{ssec:losses}.
The optimal value of $q$ in $\mathcal{L}_{q}$ for each model is determined through grid search (0.5 and 0.7 for baseline and DenSE).
For the discard approach ($\mathcal{L}_{q,discard}$), we experiment with $n_1 \in \left\lbrace 10,15,20,25\right\rbrace$ epochs, $m \in \left\lbrace 0.93,0.96,0.99\right\rbrace$, and using percentiles of loss values to discard $\left\lbrace 1,3,5\right\rbrace$ T-F patches.
Note that we are discarding T-F patches at every mini-batch, and not entire clips (see Section \ref{sec:setup}).
Results in Table~\ref{tab:results_loss} show accuracy boosts\footnote{Performance differences are expressed in terms of absolute accuracy.} of 0.4\% and 0.6\% over using plain $\mathcal{L}_{q}$, by discarding 5 patches/mini-batch and using $m=0.93$ for the baseline and DenSE models, respectively.
\begin{table}[!t]
\vspace{-3mm}
\caption{Average classification accuracy (\%) and 95\% confidence interval (7 runs) obtained by loss function approaches. $n_1$ indicates number of epochs prior to instance selection for Baseline $|$ DenSE.}
\vspace{+1mm}
\centering
\begin{tabular}{lcc}
\toprule
\textbf{Approach} & \textbf{Baseline} & \textbf{DenSE}\\
\midrule
CCE (same as Baseline in Table~\ref{tab:results})    & 66.5 $\pm$ 0.6	    & 67.9 $\pm$ 0.7     \\
$\mathcal{L}_{q}$                    	            & 68.4 $\pm$ 0.5	    & 69.2 $\pm$ 0.8       \\
$\mathcal{L}_{q,discard}$ ($n_1=25 | n_1=10$)        & 68.8 $\pm$ 0.9	    & 69.8 $\pm$ 0.7  \\
$\mathcal{L}_{q,prune}$ ($n_1=20 | n_1=15$)      & \textbf{69.0} $\pm$ 0.6	  & \textbf{70.2} $\pm$ 0.5 \\
\bottomrule
\end{tabular}
\label{tab:results_loss}
\vspace{-5mm}
\end{table}
For the pruning approach ($\mathcal{L}_{q,prune}$), rejecting clips using percentiles yields better results. 
We explore pruning $\left\lbrace 100,200,300,400,500\right\rbrace$ clips in the train set using the corresponding loss percentiles after $n_1 \in \left\lbrace 10,15,20 \right\rbrace$ epochs. 
In order to compute loss values at the clip-level, we aggregate the patch-level losses using arithmetic mean.
Improvements of 0.6\% and 1.0\% can be seen over plain $\mathcal{L}_{q}$ by pruning 200 and 400 clips from the train set, for the baseline and DenSE models, respectively.
The approach based on $\mathcal{L}_{q,prune}$ slightly outperforms the one based on $\mathcal{L}_{q,discard}$ for the two models considered.
Additionally, we observe that results yielded by $\mathcal{L}_{q,discard}$ are a bit more stochastic than those of $\mathcal{L}_{q,prune}$. 
This could happen as the former discards patches from a tiny distribution of losses (64 instances in our case), compared to considering the entire train set loss distribution in the latter.
Regarding the models, DenSE attains higher accuracy boosts with respect to plain $\mathcal{L}_{q}$ and, in general, we observe that results obtained with DenSE are more stable than with the baseline.
We hypothesize this occurs because DenSE is more accurate, thus allowing better identification of the corrupted examples.
Additionally, it has less weights, which makes it less prone to noise memorization.
These aspects make it more suitable for the proposed methods.


\vspace{-2mm}
\subsection{Discussion}
\vspace{-1mm}
\label{ssec:discussion}
The proposed methods are easy to incorporate to existing pipelines.
LSR and mixup can be added as simple functions within the data loader that feeds the network.
$\mathcal{L}_{q,discard}$ can be implemented as a custom loss function by providing information of the current epoch.
The approach based on $\mathcal{L}_{q,prune}$ can be easily added to any training procedure with few lines of code.
Furthermore, all methods cause minimal computational overhead.
The top-performing approaches on the baseline system are those based on noise-robust loss functions, especially $\mathcal{L}_{q,prune}$, which provides an accuracy increase over the CCE baseline of up to 2.5\%.
It must be noted that while we prune the dataset only once, the pruning could be done several times in an iterative fashion until convergence, potentially improving performance.
Also, this method can be used for dataset cleaning.
However, these approaches seem to be highly dependent on the period $n_1$ prior to instance selection and on the amount of rejected instances, which in turn depend on the model used, the dataset and its type and amount of label noise.
We also note that some of the reported accuracy scores feature not small confidence intervals. 
Beyond the non-deterministic nature of results obtained with GPU, we conjecture that some stochasticity is due to the noisiness of the labels.
For instance, the validation set used is composed of noisy labeled data, and we early-stop models' training by monitoring validation accuracy.
However, the fact that the model performs better on the validation set does not necessarily mean that it will perform better on the test set composed of correct labels, this being an inherent problem of dealing with noisy labels.
We leave for future work to explore more adequate evaluation strategies.
\vspace{-1mm}

\section{Conclusion}
\label{sec:conclusion}
\vspace{-1mm}
We evaluate three model-agnostic approaches to handling noisy labels when training deep networks for sound event classification.
The evaluation is carried out using the noisy set of the FSDnoisy18k dataset.
The main advantage of these methods is that they can be easily plugged into existing deep learning pipelines, requiring minimal intervention and computational overhead. 
When incorporated to the training of two different CNN architectures, these methods provide absolute accuracy boosts in the range $\approx$ 1.5 -- 2.5\%.
The proposed approaches based on noise-robust loss functions yield the highest performance, once the right parameterization is found.
We hope deep learning practitioners dealing with label noise in sound recognition can add the proposed methods into their pipelines.



\bibliographystyle{IEEEtran}
\bibliography{refs19}
%
%
%
%
%
%
%
%
%

\end{sloppy}
\end{document}